\documentclass[apsprd]{revtex4-1}
\usepackage{amsfonts}
\usepackage{amsmath}
\usepackage{amssymb}
\usepackage{graphicx}
\usepackage{color}

\def\be{\begin{equation}}
\def\te{\end{equation}}
\def\ee{\end{equation}}
\def\ba{\begin{eqnarray}}
\def\bea{\begin{eqnarray}}
\def\nn{\nonumber\\}
\def\tea{\end{eqnarray}}
\def\ea{\end{eqnarray}}
\def\eea{\end{eqnarray}}

\begin{document}

\title{Fully developed relativistic turbulence}

\author{Esteban Calzetta}
\email{calzetta@df.uba.ar}
\affiliation{Departamento de F\'isica, Facultad de Ciencias Exactas y Naturales, Universidad de Buenos Aires and IFIBA, 
CONICET, Cuidad Universitaria, Buenos Aires 1428, Argentina}


\begin{abstract}
We use a simple model consisting of energy-momentum tensor conservation and a Maxwell-Cattaneo equation for its viscous part to study nonlinear phenomena in a real relativistic fluid. We focus on new types of behavior without nonrelativistic equivalents, such as an entropy cascade driven by fluctuations in the tensor degrees of freedom of the theory. We write down the von K\'arm\'an-Howarth equations for this kind of turbulence, and consider the correlations corresponding to fully developed turbulence.
\end{abstract}
\maketitle


\section{Introduction}
Although relativistic turbulence finds many applications in astrophysical contexts \cite{BN11,LBM13,RZ13}, it is in the fields of relativistic heavy ion collisions (RHICS) \cite{H12,RR19} and reheating after inflation \cite{L19,FFTV20} that we find the whole range of dynamical possibilities in the theory. In both cases we find an scenario known as ``turbulent thermalization'' \cite{MT04,BBSV14,CH08}. Some mechanism (the original collision in RHICS, or the decay of the inflaton in reheating after inflation) brings a narrow set of modes to a highly excited state. The excitation then spreads in mode space forming a turbulent cascade, and finally the system returns to the linear regime and relaxes towards equilibrium. It is the intermediate, turbulent stage, which is the focus of attention here. It must be noted that although we are concerned with very fast processes, a hydrodynamical description is nevertheless relevant, as hydrodynamics acts as an attractor for the more complex underlying dynamics \cite{Bazow16,SND}. It is therefore important to develop an understanding of what kind of flow patterns may be supported by a turbulent, real relativistic fluid. In spite of many important contributions  \cite{MT03,GC02,K08,FW11,CR11,F13,AKLN14} such an understanding is still lacking.

Recently  \cite{ED18}, it has been proposed that relativistic turbulence proceeds not through an energy cascade, but an entropy \cite{O49} one (see also \cite{ED18b,E18,Biferale}). If confirmed, this insight will mean a substantial step forward in our understanding of the subject. Our goal is to provide a further illustration of the issues at hand by studying relativistic turbulence under the assumption of homogeneity and isotropy.

Unlike the case for ideal fluids \cite{RR13,ZR13,Lehner12,Oz15}, for real relativistic fluids there is no agreement on what are the equations of motion. There are two large families of theories, the so-called first order theories (FOTs) where the dynamical variables are the same for real and ideal fluids \cite{VanBiro12,BDN18,BDN19,BDN20,kovtun19,DasFlorNoRy20,GPRuRe19,HK20,FO10}, and the second order theories (SOTs) where the set of variables for a real fluid is larger than for an ideal one. In a typical second order theory, the viscous part of the energy momentum tensor is regarded as a hydrodynamic variable on its own. Theories of this class are Israel-Stewart \cite{israel76,IsSte76,IsSte79a,IsSte79b,IsSte80}, Divergence type Theories \cite{LiMuRu86,GerLind90,GerLind91,ReNa97,PRCal09,PRCal10,cal15,cal98,LheReRu18,lucas19,Nahuel20}, DNMR \cite{DMNR10,DMNR11,DMNR11b,DMNR12,DMNR12b,DMNR14,DMNR12c,DMNR14b,DMNR16} and Anisotropic Hydrodynamics \cite{ST14,ST14b,FMR15,FRST16,NMR17,BHS14,BNR16a,BNR16b,NBH21}. An important feature of SOTs is that they clearly display the dynamics of the traceless, divergenceless part of the energy momentum tensor. This is very relevant in the application of the theory to reheating after inflation, because this is the part of the energy momentum tensor which couples to gravitational waves \cite{LBM13,MGC17,MG21}, raising the possibility that early turbulence left an imprint on the cosmic background of gravitational radiation \cite{CF}.

Even after choosing to work within the SOT framework, a large number of alternative models remain. We shall study turbulent behavior in a minimal model where the equations of motion are energy-momentum conservation and a Maxwell-Cattaneo like equation for the viscous part thereof \cite{Max67,Catt48,Catt58,JosPrez89}. As in nonrelativistic turbulence, we shall keep only quadratic terms in the equations of motion. The motivation for this model is further discussed in Appendix (\ref{MCapp}). See also  \cite{PRC10a,PRC13a,lucas19}.

Following \cite{ED18}, our interest is in phenomena with no nonrelativistic equivalents. Because of this and its relevance to reheating after inflation we shall focus on the, maybe unrealistic, case where turbulence is driven by the tensor degrees of freedom in the theory. Our goal is to discuss what are the relevant correlation functions for the relativistic turbulent fluctuations, what constraints are placed on them by homogeneity and isotropy, what are the relevant von K\'arm\'an - Howarth equations \cite{vKH,LL87,Frisch,MY71,MY75,Chandra54,Pope}, and finally whether there is something like ``fully developed'' relativistic turbulence.

It ought to be noted that we are considering fluids very far from equilibrium. The thermal fluctuations of a relativistic real fluid are discussed in \cite{Nahuel20}.

This comment is organized as follows. We present our model in next section; as said, further motivation is given in Appendix (\ref{MCapp}). The model is such that positive entropy production is built in; nevertheless, we provide a direct check in section (\ref{entropy}), whereby we identify the relevent entropy flux, whose scale invariance will signal an entropy cascade. Then in section (\ref{isot}) we analyze the fluid correlation functions under the assumption of homogeneity and isotropy. In sections (\ref{vKH}) and (\ref{full})  we discuss the relativistic von K\'arm\'an - Howarth equations and the would be fully developed turbulence. We conclude with some final remarks.

\section{The model}
\label{model}
To investigate the nonlinear behavior of relativistic real fluids, we shall assume a minimal model, consisting of the conservation law for the energy-momentum tensor and a Maxwell-Cattaneo equation for the viscous part thereof. 

Let us analyze these currents more closely. As usual, we may decompose
\be
T^{\mu\nu}=\rho\left[u^{\mu}u^{\nu}+\frac13\Delta^{\mu\nu}+\Pi^{\mu\nu}\right],
\label{EMT}
\te
where $\rho=\rho_0e^{\delta}$ is the energy density. Here $\rho_0$ is a fiducial homogeneous background, and $\delta$ represents the inhomogeneous deviations from it. We assume the conformal equation of state $p=\rho/3$. $u^{\mu}$ is the Landau-Lifshitz fluid four velocity obeying $u^2=-1$, and
\be
\Delta_{\mu\nu}=\eta_{\mu\nu}+u_{\mu}u_{\nu}
\label{Delta}
\te
is the orthogonal projector. The viscous energy-momentum tensor $\Pi^{\mu\nu}$ is assumed to be dimensionless, transverse and traceless $\Pi^{\mu}_{\mu}=\Pi^{\mu\nu}u_{\nu}=0$.

Energy conservation leads to 
\bea
&&\dot\delta+\frac43\theta+\Pi^{\rho\sigma}u_{\rho,\sigma}=0\nn
&&\frac43\dot u^{\nu}+\frac13\Delta^{\nu\rho}\delta_{,\rho}+\Pi^{\mu\nu}_{,\mu}+\Pi^{\mu\nu}\delta_{,\mu}- u^{\nu}\Pi^{\rho\sigma}u_{\rho,\sigma}=0,
\label{conservation}
\tea
where $\dot x=u^{\mu}x_{,\mu}$, and
\be
\theta=u^{\mu}_{,\mu}.
\label{theta}
\te
The Maxwell-Cattaneo equation of motion for $\Pi^{\mu\nu}$ reads (see Appendix (\ref{MCapp}))
\bea
0&=&\dot\Pi^{\mu\nu}+\left[\frac1{\lambda}-\frac{10}{21}\theta\right]\Pi^{\mu\nu}+\frac4{15}\left[\Delta^{\rho\mu}u^{\nu}_{,\rho}+\Delta^{\rho\nu}u^{\mu}_{,\rho}-\frac23\Delta^{\nu\mu}\theta\right]\nn
&+&\frac67\left[\Pi^{\nu\rho}u^{\mu}_{,\rho}+\Pi^{\rho\mu}u^{\nu}_{,\rho}-\frac23\Delta^{\nu\mu}\Pi^{\rho\sigma}u_{\rho,\sigma}\right]+\frac14\left[u^{\nu}\Pi^{\rho\mu}+u^{\mu}\Pi^{\nu\rho}\right]\delta_{,\rho}\nn
&-&\frac17\left[\Delta^{\mu\sigma}\Pi^{\nu\rho}+\Delta^{\nu\sigma}\Pi^{\mu\rho}-\frac23\Delta^{\nu\mu}\Pi^{\rho\sigma}\right]u_{\rho,\sigma}.
\label{MC}
\tea
\subsection{Entropy}
\label{entropy}
In this model, the entropy flux
\be
S^{\mu}=S\left(\rho,x,y\right)u^{\mu},
\label{relflux}
\te
where $x=\Pi^{\mu}_{\nu}\Pi^{\nu}_{\mu}$, $y=\Pi^{\mu}_{\rho}\Pi^{\rho}_{\nu}\Pi^{\nu}_{\mu}$.
So
\be
S^{\mu}_{,\mu}=\dot S+S\theta.
\te
On dimensional grounds, 
\be
S=\rho^{3/4}s\left(x,y\right),
\label{relflux2}
\te 
so
\bea
S^{\mu}_{,\mu}&=&\rho^{3/4}\left\{s\left[\frac34\dot\delta+\theta\right]+2\dot\Pi^{\nu\rho}\Pi_{\nu\rho}\frac{\partial s}{\partial x}+3\dot\Pi^{\nu\rho}\Pi^2_{\nu\rho}\frac{\partial s}{\partial y}\right\}\nn
&=&\rho^{3/4}\left\{-\frac34\Pi^{\mu\nu}u_{\mu,\nu}s+2\dot\Pi^{\nu\rho}\Pi_{\nu\rho}\frac{\partial s}{\partial x}+3\dot\Pi^{\nu\rho}\Pi^2_{\nu\rho}\frac{\partial s}{\partial y}\right\}.
\tea
To order $\Pi_{\mu\nu}^3$, we propose an entropy density
\be
s=s_0e^{-\alpha x-\beta y};
\label{entrodensity}
\te
then, keeping only terms quadratic in $\Pi^{\mu\nu}$
\bea
S^{\mu}_{,\mu}&=&s\left\{-\frac34\Pi^{\mu\nu}u_{\mu,\nu}+2\alpha\left[\left[\frac1{\lambda}-\frac{10}{21}\theta\right]x+\frac8{15}\Pi^{\rho\sigma}u_{\rho,\sigma}+\frac{10}7\Pi^{2\mu\rho}u_{\mu,\rho}\right]\right.\nn
&+&\left.\frac8{5}\beta\left[\Pi^{2\mu\rho}u_{\mu,\rho}-\frac13\theta x\right]\right\}.
\tea
The coefficients are determined by the condition that the terms containing derivatives, which are not positive definite, must cancel. The solution is
\bea
\alpha&=&\frac{45}{64}\nn
\beta&=&-\frac{25}{14}\alpha.
\tea
In this way the entropy production is positive and reads
\be
S^{\mu}_{,\mu}=\frac{45}{32\lambda}s\;\Pi^{\mu\nu}\Pi_{\mu\nu}.
\label{entroprod}
\te

\subsection{$3+1$ Decomposition}
So far the discussion has been fully covariant, but it is convenient to adopt an explicit $3+1$ decomposition to bring forward the independent degrees of freedom in the theory. In principle this decomposition may be worked out in any frame but it is natural to choose a particular frame to simplify the dynamics as much as possible. The physical picture we have in mind is that of a fluid contained in a domain of finite size, say with a linear scale $L$. Then we may choose the frame where the fluid in globally at rest. This is a well defined concept since it is enough to measure the gravitational field sourced by the fluid, very far from the fluid itself. Of course after this the theory is no longer covariant. 

To separate explicitly time and space variables we begin by defining
\be
u^{\mu}=u^0v^{\mu},
\label{3+1}
\te
where $v^{\mu}=\left(1,{v^k}/c\right)$ and $u^0=\left(1-v^2/c^2\right)^{-1/2}$. Then
\bea
u^{\mu,\nu}&=&u^0v^{\mu,\nu}+\frac{\left(u^{0}\right)^3}{c^2}v^{\mu}v_{\lambda}v^{\lambda,\nu}=u^0\left[v^{\mu,\nu}+v^{\mu}b^{\nu}\right]\nn
\theta&=&u^0\left[v^{\lambda}_{,\lambda}+a\right],
\tea
where $b^{\nu}=\left(u^{0}\right)^2v_{\lambda}v^{\lambda,\nu}/c^2$, $a=v_{\nu}b^{\nu}$. Also call $X'=v^{\mu}X_{,\mu}$. Observe that 
$\dot u^{\nu}=u^0u'^{\nu}=\left(u^{0}\right)^2\left[v'^{\nu}+av^{\nu}\right]$ and $\left(u^{0}\right)^2v^{\mu}v_{\mu}=-1$. Also
\be
v_{\nu}\Pi^{\mu\nu}_{,\mu}+\Pi^{\rho\sigma}v_{\rho,\sigma}=0.
\te
We may decompose $\Pi^{\mu\nu}$ as
\bea
\Pi^{0i}&=&\frac{v_j}c\Pi^{ij}\nn
\Pi^{00}&=&\frac{v_iv_j}{c^2}\Pi^{ij}
\tea
and
\be
\Pi^j_j=\frac{v_iv_j}{c^2}\Pi^{ij}.
\te
This suggests writing
\be
\Pi^{ij}=\bar{\Pi}^{ij}+\frac13\delta^{ij}\Pi^k_k,
\te
$\bar\Pi^j_j=0$. Then
\be
\frac{v_iv_j}{c^2}\Pi^{ij}=\frac{v_iv_j}{c^2}\bar{\Pi}^{ij}+\frac1{3c^4}{v^2}{v_iv_j}\Pi^{ij}
\te
and
\be
\Pi^{ij}=\bar{\Pi}^{ij}+\frac1{3c^2}\delta^{ij}\frac{{v_kv_l}\bar{\Pi}^{kl}}{1-\frac13\frac{v^2}{c^2}}.
\te
We introduce these variables in the equations of motion  keeping only quadratic nonlinearities. Observe that $b^{\nu}$ is already quadratic in $v^{\mu}$ and $a$ is cubic. So we have 
\bea
0&=&c\delta'+\frac43v^{j}_{,j}+\bar\Pi^{jk}v_{j,k}\nn
0&=&\frac43cv'^{j}-\frac49v^{j}v^{k}_{,k}+\frac13c^2\delta^{,j}+\left(v_k\Pi^{jk}\right)_{,t}+c^2\bar\Pi^{jk}_{,k}+c^2\bar\Pi^{jk}\delta_{,k}\nn
0&=&c\bar\Pi'^{jk}+\left[\frac c{\lambda}-\frac{10}{21}v^{l}_{,l}\right]\bar\Pi^{jk}+\frac4{15}\left[v^{j,k}+v^{k,j}-\frac23\delta^{jk}v^{l}_{,l}\right]\nn
&+&\frac67\left[\bar\Pi^{jl}v^{k}_{,l}+\bar\Pi^{kl}v^{j}_{,l}-\frac23\delta^{jk}\bar\Pi^{lm}v_{l,m}\right]-\frac17\left[\bar\Pi^{jl}v_{l}^{,k}+\bar\Pi^{kl}v_{l}^{,j}-\frac23\delta^{jk}\bar\Pi^{lm}v_{l,m}\right]\nn
&-&\frac1{5}\left[v^{j}\left(\frac13\delta^{,k}+\bar\Pi^{kl}_{,l}\right)+v^{k}\left(\frac13\delta^{,j}+\bar\Pi^{jl}_{,l}\right)-\frac23\delta^{jk}v_m\left(\frac13\delta^{,m}+\bar\Pi^{ml}_{,l}\right)\right].
\label{nonlinear}
\tea
In the only quadratic term still containing time derivatives, we replace these derivatives by their value according to the linearized form of eqs. (\ref{nonlinear}) to get
\bea
0&=&c\delta'+\frac43v^{j}_{,j}+\bar\Pi^{jk}v_{j,k}\nn
0&=&cv'^{j}-\frac1{5}v^{j}v^{k}_{,k}+\frac1{4}c^2\delta^{,j}-\frac{3c}{4\lambda}v_k\bar\Pi^{jk}-\frac1{5}v_k\left[v^{j,k}+v^{k,j}\right]+\frac34c^2\bar\Pi^{jk}_{,k}+\frac9{16}c^2\bar\Pi^{jk}\delta_{,k}\nn
0&=&c\bar\Pi'^{jk}+\left[\frac c{\lambda}-\frac{10}{21}v^{l}_{,l}\right]\bar\Pi^{jk}+\frac4{15}\left[v^{j,k}+v^{k,j}-\frac23\delta^{jk}v^{l}_{,l}\right]\nn
&+&\frac67\left[\bar\Pi^{jl}v^{k}_{,l}+\bar\Pi^{kl}v^{j}_{,l}-\frac23\delta^{jk}\bar\Pi^{lm}v_{l,m}\right]-\frac17\left[\bar\Pi^{jl}v_{l}^{,k}+\bar\Pi^{kl}v_{l}^{,j}-\frac23\delta^{jk}\bar\Pi^{lm}v_{l,m}\right]\nn
&-&\frac1{5}\left[v^{j}\left(\frac13\delta^{,k}+\bar\Pi^{kl}_{,l}\right)+v^{k}\left(\frac13\delta^{,j}+\bar\Pi^{jl}_{,l}\right)-\frac23\delta^{jk}v_m\left(\frac13\delta^{,m}+\bar\Pi^{ml}_{,l}\right)\right].
\tea

\section{Homogeneous isotropic flows}
\label{isot}

We shall now investigate the structure of the velocity and viscous energy momentum tensor correlations in a statistically homogeneous, isotropic flow.

It is convenient to separate the equations of motion into scalar, vector and tensorial equations. For this end, we define
\bea
v_j&=&\phi_{,j}+V_j\nn
\bar\Pi_{ij}&=&\psi_{,ij}-\frac13\delta_{ij}\mathbf{\Delta}\psi+q_{i,j}+q_{j,i}+Q_{ij},
\tea
where
\be
V^j_{,j}=q^j_{,j}=Q^{jk}_{,k}=Q^j_j=0.
\te
The linearized equations of motion decouple, so we have a theory with scalar degrees of freedom $\left(\delta,\phi,\psi\right)$, vector degrees of freedom $\left(V_j,q_j\right)$ and a tensorial degree of freedom $Q$. These modes propagate at different velocities \cite{Muller99,BL99}. For our model we get $\sqrt{3/5}c$ for the scalar modes, $\sqrt{1/5}c$ for vector modes, and $0$ for tensor modes. In other words, the tensor sector are the ``slow'' degrees of freedom, and there is an incompressible limit where scalar modes are frozen, and vector modes are slaved to tensor ones. We shall work within this regime in what follows, namely, we shall disregard the scalar degrees of freedom and assume that $\bar\Pi^{jk}$ contains tensor degrees of freedom only (since it then becomes identical with $Q^{jk}$, we shall drop this latter notation). Then the EOMs become
\bea
0&=&v'^{j}-\frac{3c}{4\lambda}v_k\bar\Pi^{jk}-\frac1{5}v_k\left[v^{j,k}+v^{k,j}\right]\nn
0&=&\bar\Pi'^{jk}+\frac c{\lambda}\bar\Pi^{jk}+\frac67\left[\bar\Pi^{jl}v^{k}_{,l}+\bar\Pi^{kl}v^{j}_{,l}-\frac23\delta^{jk}\bar\Pi^{lm}v_{l,m}\right]\nn
&-&\frac17\left[\bar\Pi^{jl}v_{l}^{,k}+\bar\Pi^{kl}v_{l}^{,j}-\frac23\delta^{jk}\bar\Pi^{lm}v_{l,m}\right].
\label{finalform}
\tea
The correlation functions most relevant to our discussion are the two point funtion
\be
\left\langle \bar\Pi^{ij}\left(x,t\right)\bar\Pi^{kl}\left(y,t\right)\right\rangle=\int\frac{d^3p}{\left(2\pi\right)^3}\;e^{ip\left(x-y\right)}F^{ijkl}\left(p,t\right)
\te
and
\be
\left\langle \bar\Pi^{il}\left(x,t\right)\bar\Pi^{jl}\left(y,t\right)v^k\left(z,t\right)\right\rangle=\int \frac{d^3pd^3q}{\left(2\pi\right)^6}\;e^{i\left[p\left(x-z\right)+q\left(y-z\right)\right]}G^{ijk}\left(p,q,t\right),
\te
where the brackets $\left\langle \right\rangle$ denote average over flow realizations. Both are severely restricted by the isotropy assumption. $F^{ijkl}\left(p,t\right)$ has to be symmetric in $\left(i,j\right)$, $\left(k,l\right)$ and under the exchange of these pairs by one another. Also $p_iF^{ijkl}=F^{jjkl}=0$. So we must have
\be
F^{ijkl}\left(p,t\right)=\frac12\left[h_p^{ik}h_p^{jl}+h_p^{il}h_p^{jk}-h_p^{ij}h_p^{kl}\right]f\left(p,t\right),
\te
where $p=\left|p\right|$ and
\be
h_p^{ik}=\delta^{ik}-\frac{p^ip^k}{p^2}
\te
Observe that $f\left(p,t\right)$ may be obtained as the Fourier transform of the fully contracted correlation
\be
\left\langle \bar\Pi^{ij}\left(x,t\right)\bar\Pi^{ij}\left(y,t\right)\right\rangle=2\int\frac{d^3p}{\left(2\pi\right)^3}\;e^{ip\left(x-y\right)}f\left(p,t\right).
\label{contracted}
\te
$G^{ijk}$ obeys $p_iG^{ijk}=q_jG^{ijk}=\left(p+q\right)_kG^{ijk}=0$ and $G^{ijk}\left(p,q,t\right)=G^{jik}\left(q,p,t\right)$. Therefore
\be
G^{ijk}\left(p,q,t\right)=\left(-i\right)h_p^{il}h_q^{jm}h_{\left(p+q\right)}^{kn}g^{lmn}\left(p,q,t\right),
\label{g31}
\te
where
\be
g^{lmn}\left(p,q,r\right)=g_1q^lp^m\left(p-q\right)^n+g_2q^l\delta^{mn}+g_2^Tp^m\delta^{ln}+g_3\left(p-q\right)^n\delta^{lm}.
\label{g32}
\te
The coefficients are functions of $\left|p\right|$, $\left|q\right|$ and $pq=p_iq^i$, and $g_2^T\left(p,q,pq,t\right)=g_2\left(q,p,pq,t\right)$. Observe that $g_1=-g_1^T$ and $g_3=-g_3^T$. We observe the contractions
\bea
iG^{jjk}&=&h_{\left(p+q\right)}^{kn}\left(p-q\right)_n\left[-g_1\left(pq\right)\left(1-\frac{\left(pq\right)^2}{p^2q^2}\right)-\frac12g_2\frac{\left(pq\right)}{p^2}\left(1+\frac{\left(pq\right)}{q^2}\right)\right.\nn
&+&\left.\frac12g_2^T\frac{\left(pq\right)}{q^2}\left(1+\frac{\left(pq\right)}{p^2}\right)+g_3\left(1+\frac{\left(pq\right)^2}{p^2q^2}\right)\right]\nn
iG^{kjk}&=&h_q^{jm}p_m\left[-\frac12g_1\Delta\left(1+\frac{\left(pq\right)}{p^2}\right)-\frac12g_2\frac{q^2}{\left(p+q\right)^2}\left(1+\frac{\left(pq\right)}{p^2}\right)\left(1+\frac{\left(pq\right)}{q^2}\right)\right.\nn
&+&\left.2g_2^T\left(1-\frac{\Delta}{8p^2}\right)+2g_3\frac{\left(pq\right)}{\left(p+q\right)^2}\left(1+\frac{\left(pq\right)}{p^2}\right)\right]\nn
iG^{ijj}&=&h_p^{il}q_l\left[\frac12g_1\Delta\left(1+\frac{\left(pq\right)}{q^2}\right)+2g_2\left(1-\frac{\Delta}{8q^2}\right)\right.\nn
&-&\left.\frac12g_2^T\frac{p^2}{\left(p+q\right)^2}\left(1+\frac{\left(pq\right)}{p^2}\right)\left(1+\frac{\left(pq\right)}{q^2}\right)-2g_3\frac{\left(pq\right)}{\left(p+q\right)^2}\left(1+\frac{\left(pq\right)}{q^2}\right)\right],
\tea
where
\be
\Delta=\left(p-q\right)^kh_{\left(p+q\right)}^{kn}\left(p-q\right)^n=\frac{4p^2q^2}{\left(p+q\right)^2}\left(1-\frac{\left(pq\right)^2}{p^2q^2}\right).
\te

\section{von K\'arm\'an-Howarth equations}
\label{vKH}
The above correlations are further constrained by the relativistic von K\'arm\'an-Howarth equations. To derive this equation, we consider eq. (\ref{finalform}) for $\bar\Pi^{jk}\left(z\right)$, fully contracted with $\bar\Pi^{jk}\left(x\right)$ (by eq. (\ref{contracted}), the full correlation may be reconstructed from the contracted form). We repeat exchanging $z$ and $x$, average over flow realizations and add to obtain
\bea
&&-\left[\frac d{dt}+\frac {2c}{\lambda}\right]\left\langle \bar\Pi^{jk}\left(x,t\right)\bar\Pi^{jk}\left(z,t\right)\right\rangle=\left\langle \bar\Pi^{jk}\left(x,t\right)\left(v^l\bar\Pi^{jk}_{,l}\right)\left(z,t\right)\right\rangle\nn
&+&\left\langle \bar\Pi^{jk}\left(z,t\right)\left(v^l\bar\Pi^{jk}_{,l}\right)\left(x,t\right)\right\rangle+\frac{12}7\left\langle \bar\Pi^{jk}\left(x,t\right)\left(\bar\Pi^{jl}v^{k}_{,l}\right)\left(z,t\right)\right\rangle+\frac{12}7\left\langle \bar\Pi^{jk}\left(z,t\right)\left(\bar\Pi^{jl}v^{k}_{,l}\right)\left(x,t\right)\right\rangle\nn
&-&\frac27\left\langle \bar\Pi^{jk}\left(x,t\right)\left(\bar\Pi^{jl}v_{l}^{,k}\right)\left(z,t\right)\right\rangle-\frac27\left\langle \bar\Pi^{jk}\left(z,t\right)\left(\bar\Pi^{jl}v_{l}^{,k}\right)\left(x,t\right)\right\rangle.
\tea
Now
\be
\left\langle \bar\Pi^{jk}\left(x,t\right)\left(v^l\bar\Pi^{jk}_{,l}\right)\left(z,t\right)\right\rangle=\left(-i\right)\int \frac{d^3p}{\left(2\pi\right)^3}\;e^{ip\left(x-z\right)}p_k\int\frac{d^3q}{\left(2\pi\right)^3}\;G^{jjk}\left(p,q,t\right).
\label{jjk}
\te
It is clearly symmetric in $\left(x,z\right)$. Similarly
\be
\left\langle \bar\Pi^{jk}\left(x,t\right)\left(\bar\Pi^{jl}v^{k}_{,l}\right)\left(z,t\right)\right\rangle=\left(-i\right)\int \frac{d^3p}{\left(2\pi\right)^3}\;e^{ip\left(x-z\right)}p_k\int\frac{d^3q}{\left(2\pi\right)^3}\;G^{jkj}\left(p,q,t\right)
\te
\be
\left\langle \bar\Pi^{jk}\left(x,t\right)\left(\bar\Pi^{jl}v^{l}_{,k}\right)\left(z,t\right)\right\rangle=\left(-i\right)\int \frac{d^3p}{\left(2\pi\right)^3}\;e^{ip\left(x-z\right)}\int\frac{d^3q}{\left(2\pi\right)^3}\;q_kG^{kjj}\left(p,q,t\right),
\te
so the von K\'arm\'an-Howarth equation reads
\bea
&-&\frac12\left[\frac d{dt}+\frac {2c}{\lambda}\right]\left\langle \bar\Pi^{jk}\left(x,t\right)\bar\Pi^{jk}\left(z,t\right)\right\rangle=\left\langle \bar\Pi^{jk}\left(x,t\right)\left(v^l\bar\Pi^{jk}_{,l}\right)\left(z,t\right)\right\rangle\nn
&+&\frac{12}7\left\langle \bar\Pi^{jk}\left(x,t\right)\left(\bar\Pi^{jl}v^{k}_{,l}\right)\left(z,t\right)\right\rangle-\frac27\left\langle \bar\Pi^{jk}\left(x,t\right)\left(\bar\Pi^{jl}v_{l,k}\right)\left(z,t\right)\right\rangle,
\label{vkh}
\tea
or in momentum domain 
\be
\frac12\left[\frac d{dt}+\frac {2c}{\lambda}\right]f\left(p,t\right)=i\int\frac{d^3q}{\left(2\pi\right)^3}\;\left[p_kG^{jjk}\left(p,q,t\right)+\frac{12}7p_kG^{jkj}\left(p,q,t\right)-\frac27q_kG^{kjj}\left(p,q,t\right)\right].
\te
In terms of the decomposition eqs. (\ref{g31}) and (\ref{g32})
\bea
ip_kG^{jjk}&=&\frac14\frac{\Delta}{p^2q^2}\left[-2g_1\left(pq\right)\left(p^2q^2-\left(pq\right)^2\right)+2g_3\left(p^2q^2+\left(pq\right)^2\right)-g_2\left(pq\right)\left(q^2+pq\right)+g_2^T\left(pq\right)\left(p^2+pq\right)\right]\nn
ip_kG^{jkj}&=&-\frac14\frac{\Delta}{p^2q^2}\left[2g_1\left(p^2q^2-\left(pq\right)^2\right)\left[p^2+pq\right]+g_2\left[q^2p^2+\left(pq\right)p^2+\left(pq\right)q^2+\left(pq\right)^2\right]\right.\nn
&-&\left.g_2^T\left(2p^4+p^2q^2+4p^2\left(pq\right)+\left(pq\right)^2\right)-2g_3\left(pq\right)\left[p^2+\left(pq\right)\right]\right]\nn
iq_kG^{kjj}&=&\frac14\frac{\Delta}{p^2q^2}\left[2g_1\left(p^2q^2-\left(pq\right)^2\right)\left(q^2+pq\right)+g_2\left(p^2q^2+2q^4+4q^2\left(pq\right)+\left(pq\right)^2\right)\right.\nn
&-&\left.g_2^T\left(q^2p^2+\left(pq\right)p^2+\left(pq\right)q^2+\left(pq\right)^2\right)-2g_3\left(pq\right)\left[q^2+pq\right]\right],
\label{jjk2}
\tea
where $\left(pq\right)=p_iq^i$

\section{Fully developed turbulence}
\label{full}
From now on we shall work in the incompressible limit where the scalar degrees of freedom are frozen. We shall also assume that $u^0\approx 1$. From the expression eqs. (\ref{relflux}), (\ref{relflux2}), (\ref{entrodensity}) and (\ref{3+1}) for the entropy density, we see that the transport term in the entropy balance equation eq. (\ref{entroprod}), to lowest nontrivial order in $\bar\Pi^{\mu\nu}$, and averaged over flow realizations, is given by
\be
cS^k_{,k}=\left(u^0v^ks\right)_{,k}\approx 2s\left[\frac{1}{c^2}\epsilon+\frac{\alpha}{\tau}\right],
\label{transport}
\te
where
\bea
\epsilon &=&\frac12\left\langle v^k\left(v^jv_{j,k}\right)\right\rangle\nn
\tau^{-1}&=&-\left\langle v^k\left(\bar\Pi^{ij}\bar\Pi_{ij,k}\right)\right\rangle,
\label{tau}
\tea
where as above the brackets $\left\langle \right\rangle$ denote average over flow realizations.
$\epsilon$ corresponds to the nonrelativistic cascade; although one may use these equations to work out the relativistic corrections to the Kolmogorov spectrum, we shall not discuss it further. On the other hand, $\tau^{-1}$ corresponds to a new kind of cascade with no nonrelativistic analog. 

The expressions eq. (\ref{tau}) require to be interpreted, since in an homogeneous flow all expectation values at a single point are traslation invariant. We take them to mean
\bea
\epsilon&\approx&\left\langle v^j\left(x\right)\left(v^kv_j\right)_{,k}\left(z\right)\right\rangle\nn
\tau^{-1}&=&-\left\langle \bar\Pi^{ij}\left(x\right)\left(v^k\bar\Pi_{ij}\right)_{,k}\left(z\right)\right\rangle=\left\langle \bar\Pi^{ij}_{,k}\left(x\right)\left(v^k\bar\Pi_{ij}\right)\left(z\right)\right\rangle.
\ref{taub}
\tea
We shall discuss later what range of $x$ and $z$ is intended.

Fully developed turbulence is the case where the dynamically generated dimensionful variable $\tau$ defined in eq. (\ref{tau}) is scale independent, 
\be
\frac{\partial}{\partial x^k}\left\langle\bar\Pi^{ij}\left(x\right)\left(v^k\bar\Pi_{ij}\right)\left(z\right)\right\rangle =\tau^{-1}=\;\mathrm{constant},
\label{tau2}
\te
or else
\be
ip_k\int\frac{d^3q}{\left(2\pi\right)^3}\;G^{jjk}\left(p,q,t\right)=\tau^{-1}\left(2\pi\right)^3\delta\left(p\right).
\label{tau3}
\te
These equations are the relativistic counterpart to the Kolmogorov $4/5$ law \cite{LL87,Frisch,MY71,MY75,Chandra54,Pope}. 

Since $\tau$ has dimensions of time, the simplest dimensionally correct ansatz for $\left\langle \bar\Pi^{ij}\left(x\right)\bar\Pi^{kl}\left(y\right)v^m\left(z\right)\right\rangle$ would be linear in $r=x-z$ and $s=y-z$. Imposing the relevant symmetry, tracelessness and divergencelessness constraints, we obtain the most general form
\be
\left\langle \bar\Pi^{ij}\left(x\right)\bar\Pi^{kl}\left(y\right)v^m\left(z\right)\right\rangle=\frac A{\tau}\left\{G^{\left(ij\right),\left(kl\right),m}\left(r\right)+
G^{\left(kl\right),\left(ij\right),m}\left(s\right)\right\},
\label{ansatz1}
\te
where
\bea
&&G^{\left(ij\right),\left(kl\right),m}\left(r\right)=r^i\left[\delta^{jk}\delta^{lm}+\delta^{jl}\delta^{km}-6\delta^{jm}\delta^{kl}\right]\nn
&+&r^j\left[\delta^{ik}\delta^{lm}+\delta^{il}\delta^{km}-6\delta^{im}\delta^{kl}\right]+r^k\left[8\left(\delta^{il}\delta^{jm}+\delta^{jl}\delta^{im}\right)-6\delta^{ij}\delta^{lm}\right]\nn
&+&r^l\left[8\left(\delta^{ik}\delta^{jm}+\delta^{im}\delta^{jk}\right)-6\delta^{ij}\delta^{km}\right]+r^m\left[-6\left(\delta^{ik}\delta^{jl}+\delta^{il}\delta^{jk}\right)+8\delta^{ij}\delta^{kl}\right],
\label{ansatz2}
\tea
which reduces to
\be
\left\langle \bar\Pi^{il}\left(x\right)\bar\Pi^{jl}\left(y\right)v^k\left(z\right)\right\rangle=\frac {7A}{\tau}
\left\{-r^i\delta^{jk}-r^k\delta^{ij}+4r^j\delta^{ik}-s^j\delta^{ik}-s^k\delta^{ij}+4s^i\delta^{jk}\right\}.
\label{ansatz3}
\te
Fourier transforming, we obtain a representation eqs. (\ref{g31}) and (\ref{g32}) with
\bea
g_1&=&0\nn
g_2&=&\left[\left(2\pi\right)^6\frac {7A}{\tau}\right]\frac {4q_r}{q^2}\frac{\partial}{\partial q_r}\delta^{\left(3\right)}\left(q\right)\delta^{\left(3\right)}\left(p\right)\nn
g_3&=&\left[\left(2\pi\right)^6\frac {7A}{\tau}\right]\frac12\left[\frac {q_r}{q^2}\frac{\partial}{\partial q_r}\delta^{\left(3\right)}\left(q\right)\delta^{\left(3\right)}\left(p\right)-\frac {p_r}{p^2}\frac{\partial}{\partial p_r}\delta^{\left(3\right)}\left(p\right)\delta^{\left(3\right)}\left(q\right)\right].
\label{ansatz4}
\tea
We therefore find
\be
\left\langle \bar\Pi^{jk}\left(x,t\right)\left(\bar\Pi^{jl}v^{k}_{,l}\right)\left(z,t\right)\right\rangle=\left\langle \bar\Pi^{jk}\left(x,t\right)\left(\bar\Pi^{jl}v^{l}_{,k}\right)\left(z,t\right)\right\rangle=
-30\frac {7A}{\tau},
\te
but
\be
\left\langle \bar\Pi^{jk}\left(x,t\right)\left(v^l\bar\Pi^{jk}_{,l}\right)\left(z,t\right)\right\rangle=0.
\te
There is a similar situation in nonrelativistic incompressible turbulence, because K41 theory requires $\left\langle v^j\left(x+r\right)\left(v_jv^k\right)\left(x\right)\right\rangle\propto\epsilon r^k$, but an actual computation shows that $\left\langle v^j\left(x+r\right)\left(v_jv^k\right)\left(x\right)\right\rangle=O\left(r^3\right)$ when $r\to 0$ \cite{Pope}.

This shows that the ansatz eq. (\ref{ansatz3}) is too naive and must be modified, and eq. (\ref{ansatz4}) suggests the modification must be to include a non zero $g_1$. By way of example, we shall discuss a simple choice for $g_1$ which is consistent with all the constraints above.

From eqs. (\ref{tau2}, \ref{tau3}) and (\ref{jjk2}), we see that what is required is
\be
\int\frac{d^3q}{\left(2\pi\right)^3}\;\frac{p^2q^2}{\left(p+q\right)^2}\left(1-\frac{\left(pq\right)^2}{p^2q^2}\right)^2g_1\left(pq\right)\stackrel{?}{=}-\frac{1}2\tau^{-1}\left(2\pi\right)^3\delta^{\left(3\right)}\left(p\right).
\label{conj}
\te
A simple way to achieve this is (see Appendix (\ref{B}))
\bea
&&g_1=-B\tau^{-1}\frac{\left(p+q\right)^2\left(pq\right)}{p^4q^4}\frac{\delta^{\left(3\right)}\left(p\right)}{q^3}\nn
&&B=\frac{105}{16}\left(2\pi\right)^6\Lambda\nn
&&\Lambda=\left[\int\frac{d^3q}{q^3}\right]^{-1}.
\label{ansatz}
\tea
The power of $q$ is chosen to give $g_1$ the right dimensions. The integral defining $\Lambda$ requires regularization. We assume there is an outer length scale $L$, for example, the size of the domain containing the fluid, and an inner scale $\ell$ (see eq. (\ref{ell}) below), so
\be
\Lambda^{-1}=\int_{L^{-1}<p<\ell^{-1}}\frac{d^3q}{q^3}=4\pi\;\ln\frac{L}{\ell}
\te
Ansatz (\ref{ansatz}) breaks the condition that $g_1=-g_1^T$, so we write instead
\be
g_1=-B\tau^{-1}\frac{\left(p+q\right)^2\left(pq\right)}{p^4q^4}\left[\frac{\delta^{\left(3\right)}\left(p\right)}{q^3}-\frac{\delta^{\left(3\right)}\left(q\right)}{p^3}\right]\nn
\te
and we get
\be
\int\frac{d^3q}{\left(2\pi\right)^3}\;\frac{p^2q^2}{\left(p+q\right)^2}\left(1-\frac{\left(pq\right)^2}{p^2q^2}\right)^2g_1\left(pq\right)=\frac{1}2\tau^{-1}\left(2\pi\right)^3\left[\frac{\Lambda}{p^3}-\delta^{\left(3\right)}\left(p\right)\right]\nn
\label{conj2}
\te
so
\be
\frac{\partial}{\partial x^k}\left\langle\bar\Pi^{ij}\left(x\right)\left(v^k\bar\Pi_{ij}\right)\left(z\right)\right\rangle =\tau^{-1}\left[\Lambda\int d^3p\;\frac{e^{i\vec p\vec r}}{p^3}-1\right]
\label{tau4}
\te
As in the nonrelativistic theory, we get the desired  behavior at not too short distances.

Let us now consider how the new $g_1$ affects the von K\'arm\'an-Howarth equation (\ref{vkh}).
Because $\left(pq\right)\Delta g_1$ is an even function of $q$, the contribution  from $g_1$ to $\left\langle \bar\Pi^{jk}\left(x,t\right)\left(\bar\Pi^{jl}v^{k}_{,l}\right)\left(z,t\right)\right\rangle$ is the same as to $\left\langle \bar\Pi^{jk}\left(x,t\right)\left(v^l\bar\Pi^{jk}_{,l}\right)\left(z,t\right)\right\rangle$, and the same holds for $ \left\langle \bar\Pi^{jk}\left(x,t\right)\left(\bar\Pi^{jl}v^{l}_{,k}\right)\left(z,t\right)\right\rangle$, but for the sign. So if $g_1$ is chosen in this way, leaving $g_2$ and $g_3$ as in eq. (\ref{ansatz4}), eq. (\ref{vkh}) becomes
\be
\left\langle \bar\Pi^{jk}\left(x,t\right)\bar\Pi^{jk}\left(z,t\right)\right\rangle\approx \frac{3\lambda}{c\tau}\left[100A-1+\Lambda\int d^3p\;\frac{e^{i\vec p\vec r}}{p^3}\right],
\label{spectrum}
\te
which corresponds to a scale invariant spectrum $\propto p^{-3}$ for $p\not=0$. Imposing the boundary condition that $\left\langle \bar\Pi^{jk}\left(x,t\right)\bar\Pi^{jk}\left(z,t\right)\right\rangle\to 0$ for large $r$ we find $A=1/100$.

Eq. (\ref{spectrum}) shows that the typical size of tensor fluctuations at a scale $r$ is $\left(\lambda/c\tau\right)\left(r/L\right)^3$. This suggests defining $\ell$ as the scale where typical fluctuations are of order $1$, namely 
\be
\ell\approx L\left(c\tau/\lambda\right)^{1/3}.
\label{ell}
\te

\section{Final Remarks} 

In this comment we have intended to show that the richer dynamics of real relativistic fluids vis a vis their nonrelativistic counterparts allows for new kinds of turbulent phenomena, and have provided an example of how turbulence driven by tensor degrees of freedom may look like in the simple case of homogeneous, isotropic, fully developed turbulence. 

The solution we obtained depended on the existence of two time scales in the theory. One is the time scale $\lambda/c$, where $\lambda$ has the meaning of a mean free path and it is carried over from the Boltzmann equation in the relaxation time approximation eq. (\ref{BE}). This scale determines the size of the entropy production in the entropy balance equation (\ref{entroprod}). The other is the scale $\tau$ introduced in eqs. (\ref{tau}) and (\ref{tau2}). Unlike $\lambda/c$, $\tau$ is not a parameter of the model, but rather of a particular statistical realization. The ratio $\lambda/c\tau$ regulates the size of tensor fluctuations and in this sense it behaves as the ``Reynold's number'' of the theory. Tensor turbulence appears when this ratio is larger than one.

The solution we have obtained is remarkable because energy-momentum conservation is upholded throughout, in the sense that there is no right hand side to equations (\ref{conservation}). The fluid is not stirred by an external ``force'', instead eddies of size $L$ produce entropy and then this entropy is distributed across scales by the transport term in equation (\ref{entroprod}). Eventually entropy is ejected into eddies of size $r\le\ell$, which registers as heat loss from the fluid, so in more complete solution we should be able to see the decay of the turbulent flow pattern.

As in the usual Kolmogorov theory, the scale free spectrum $\propto p^{-3}$ could have been derived on dimensional grounds alone. However, it is interesting to see that, again as in the usual Kolmogorov theory, two and three point correlations are closely related through the respective von K\'arm\'an-Howarth equations. Another point of contact is that the presence of three point correlations indicates a strong departure from gaussianity. 

Of course the example we have presented is built on very limiting premises, namely the extreme parameter range $\tau\le \lambda/c$, the exclusion of scalar degrees of freedom, and the subordination of vector degrees of freedom to tensor ones. Clearly a more complete solution is required before we can discuss its relevance to actual observations. However, we believe our work constitutes a proof of principle. 

It also must be kept in mind that, as we mentioned in the Introduction, there are several approaches to relativistic real hydrodynamics. Even if we restrict ourselves to SOTs, these different approaches contain different nonlinear terms and different free parameters, therefore the resulting von K\'arm\'an-Howarth equations must be expected to be different as well. In this sense our model, which has the mean free path $\lambda$ as the only adjustable parameter, may be regarded as a minimal model which still enforces nonnegative entropy production to all orders in departures from equilibrium. We believe the patterns we have discussed will still be found in other approaches such as DNMR \cite{DMNR10,DMNR11,DMNR11b,DMNR12,DMNR12b,DMNR14,DMNR12c,DMNR14b,DMNR16} and Anisotropic Hydrodynamics 
\cite{ST14,ST14b,FMR15,FRST16,NMR17,BHS14,BNR16a,BNR16b,NBH21}, but defer detailed comparison to future work.

To conclude, we regard this work as a proof of principle that real relativistic fluids may display flow patterns which are qualitatively different from those present in nonrelativistic hydrodynamics, but that nevertheless some insight may be gained by using tools already well developed in the turbulence literature. Even if it would be hard to reproduce these patterns in actual situations, their existance may influence observable flows, not unlike a nonthermal fixed point structures flow in nonequilibrium field theory even if not directly approached \cite{BRS08,Berges15,NEKSSG12,SMG19}.

Our long term goal is to assemble a turbulence toolkit to assist in the development of  phenomenological models of  reheating after inflation \cite{L19}. This is certainly the most violent and far from equilibrium process in the history of the Universe. As suggested some time ago \cite{GC02,MT03,MT04}, turbulence may provide a qualitative understanding of this process, which complements full scale numerical simulations already under way \cite{FFTV20}.

\appendix
\section{Motivating eq. (\ref{MC})}
\label{MCapp}
We consider the hydrodynamical description for a gas of massless particles obeying a Boltzmann equation with an Anderson-Witting collision term \cite{RZ13,AW74a,AW74b}
\be
p^{\mu}f_{\mu}=\frac1{\lambda}{u^{\mu}p_{\mu}}\left[f-f_0\right],
\label{BE}
\te
where $f_0$ is an equilibrium 1pdf
\be
f_0=e^{\beta_0^{\mu}p_{\mu}}
\te
and
\be
u_{\mu}\left[T^{\mu\nu}-T_0^{\mu\nu}\right]=0,
\te
thus enforcing EMT conservation. 

Our primary concern is that the hydrodynamic description should enforce energy-momentum conservation and positive entropy production. One simple way to achieve the second requirement is to postulate an ansatz \cite{lucas19}
\be
f=e^{\sum_A\zeta_A\varphi^A\left(p,x\right)},
\te
where the $\varphi^A\left(p,x\right)$ are known functions of position and momentum, and the $\zeta_A$ will be the hydrodynamic variables. The equations of motion are obtained from the moments of the kinetic equation
\be
\int \frac{2d^4p}{\left(2\pi\right)^3}\;\delta\left(-p^2\right)\theta\left(p^0\right)\;\varphi^A\left(p,x\right)\left\{p^{\mu}f_{\mu}-\frac1{\lambda}{u^{\mu}p_{\mu}}\left[f-f_0\right]\right\}=0.
\te
Positive entropy production follows from the kinetic theory $H$ theorem. The equations of motion are conservation laws for currrents 
\be
A^{A\mu}=\int \frac{2d^4p}{\left(2\pi\right)^3}\;\delta\left(-p^2\right)\theta\left(p^0\right)\;\varphi^A\left(p,x\right)p^{\mu}f_{\mu}.
\te
By choosing $\varphi^0=p^{\mu}$, $\zeta_0=\beta_{\mu}$ we enforce energy momentum conservation as one of the equations of motion. Our second choice will be $\varphi^{1}=p^{\mu}p^{\nu}/\left(-U_{\rho}p^{\rho}\right)$. $U^{\mu}$ is a fiducial unit vector to be identified with $u^{\mu}$ after deriving the equations of motion, and we assume $\zeta^{\mu}_{\mu}=U^{\mu}\zeta_{\mu\nu}=0$ for a conformal fluid. This choice has been very succesful in reproducing flows in Bjorken and Gubser backgrounds \cite{lucas19}. We get the currents
\bea
T^{\mu\nu}&=&\frac{\partial\Phi^{\mu}}{\partial\beta_{\nu}}=\frac 2{\left(2\pi\right)^3}\int\;d^4p\;\delta\left(-p^2\right)\theta\left(p^0\right)\;p^{\mu}p^{\nu}f\nn
A^{\mu\nu\rho}&=&\frac{\partial\Phi^{\mu}}{\partial\zeta_{\nu\rho}}=\frac 2{\left(2\pi\right)^3}\int\;d^4p\;\delta\left(-p^2\right)\theta\left(p^0\right)\;p^{\mu}\frac{p^{\nu}p^{\rho}}{\left(-U_{\theta}p^{\theta}\right)}f\nn
S^{\mu}&=&\frac 2{\left(2\pi\right)^3}\int\;d^4p\;\delta\left(-p^2\right)\theta\left(p^0\right)\;p^{\mu}\left[1-\ln f\right]f.
\tea
$T^{\mu\nu}$ and $A^{\mu\nu\rho}$ are symmetric and traceless on any pair of indexes, and $u_{\mu}A^{\mu\nu\rho}=-T^{\nu\rho}$. 
The fluid equations are 
\bea
T^{\mu\nu}_{,\nu}&=&0\nn
S_{\lambda\chi\nu\rho}\left[A^{\mu\nu\rho}_{,\mu}-K^{\mu\nu\rho\sigma}U_{\sigma,\mu}+\frac1{\lambda}T^{\nu\rho}\right]&=&0,
\tea
where
\be
S_{\lambda\chi\nu\rho}=\frac12\left[\Delta_{\lambda\nu}\Delta_{\lambda\rho}+\Delta_{\lambda\rho}\Delta_{\lambda\nu}-\frac23\Delta_{\lambda\chi}\Delta_{\nu\rho}\right]
\te
($\Delta_{\mu\nu}$ as in eq. (\ref{Delta})) and
\be
K^{\mu\nu\rho\sigma}=\frac 2{\left(2\pi\right)^3}\int\;d^4p\;\delta\left(-p^2\right)\theta\left(p^0\right)\;p^{\mu}\frac{p^{\nu}p^{\rho}p^{\sigma}}{\left(-U_{\theta}p^{\theta}\right)^2}f.
\te
The fluid equations imply
\be
S^{\mu}_{,\mu}=\frac1{\lambda}\zeta_{\nu\rho}T^{\nu\rho}
\te
and so the Second Law is enforced as long as 
\be
\zeta_{\nu\rho}T^{\nu\rho}\ge 0.
\te
Although we know that positive entropy production follows from the kinetic theory $H$ theorem, we may also go through a direct check. In a frame where $\zeta_{ij}=\mathrm{diag}\left(\zeta_++\zeta_-,\zeta_+-\zeta_-,-2\zeta_+\right)$ and writing 

\be
p^i=p\left(\sin\theta\;\cos\varphi, \sin\theta\;\sin\varphi,\cos\theta\right),
\te 
we have
\be
\zeta_{\nu\rho}T^{\nu\rho}=\vec{\zeta}\cdot\nabla J\left(\vec{\zeta}\right)
\te
where $\vec{\zeta}=\left(\zeta_+,\zeta_-\right)$ and
\be
J=\int\frac{p^2dp}{\left(2\pi\right)^3}e^{-p/T}\int_0^{\pi}d\theta\;\sin\theta\;e^{p\zeta_+\left(1-3\cos^2\theta\right)}\int_{-\pi}^{\pi}d\varphi\;e^{p\zeta_-\sin^2\theta\cos 2\varphi}.
\te
If $\vec\zeta =\zeta\hat{\zeta}$, $\hat{\zeta}^2=1$, then
\be
\zeta_{\nu\rho}T^{\nu\rho}=\zeta\frac {d}{d\zeta}J\left(\zeta\hat{\zeta}\right),
\te
and the positivity follows from
\bea
\frac {d}{d\zeta}J\left(0\right)&=&0\nn
\frac {d^2}{d\zeta^2}J\left(\zeta\hat{\zeta}\right)&\ge& 0.
\tea
Let us analyze the hydrodynamic currents more closely. As usual, we may decompose the energy-momentum tensor as in eq. (\ref{EMT}). Then, the requirement that
\be
u_{\mu}A^{\mu\nu\rho}=-T^{\nu\rho}
\te
implies
\be
A^{\mu\nu\rho}=\rho\left\{u^{\mu}u^{\nu}u^{\rho}+\frac13\left[u^{\mu}\Delta^{\nu\rho}+u^{\nu}\Delta^{\mu\rho}+u^{\rho}\Delta^{\nu\mu}\right]+u^{\mu}\Pi^{\nu\rho}+u^{\nu}\Pi^{\mu\rho}+u^{\rho}\Pi^{\nu\mu}\right\}.
\te
Energy conservation leads to eqs. (\ref{conservation}), and
\bea
S^{\nu\rho\lambda\chi}A^{\mu}_{\lambda\chi,\mu}&=&\rho\left\{\dot\Pi^{\nu\rho}-\frac13\theta\Pi^{\nu\rho}+\frac13\left[\Delta^{\mu\rho}u^{\nu}_{,\mu}+\Delta^{\mu\nu}u^{\rho}_{,\mu}-\frac23\Delta^{\nu\rho}\theta\right]\right.\nn
&+&\left.\Pi^{\nu\mu}u^{\rho}_{,\mu}+\Pi^{\rho\mu}u^{\nu}_{,\mu}-\frac23\Delta^{\nu\rho}\Pi^{\mu\sigma}u_{\mu,\sigma}+\frac14\left[u^{\nu}\Pi^{\rho\mu}+u^{\rho}\Pi^{\nu\mu}\right]\delta_{,\mu}\right\},
\tea
where we have used the equations for $\dot\rho$ and $\dot u^{\nu}$ disregarding quadratic terms in $\Pi^{\mu\nu}$. 

It only remains to
compute $K^{\mu\nu\rho\sigma}$ up to linear terms in $\Pi^{\mu\nu}$. Let us write the generic form
\bea
K^{\mu\nu\rho\sigma}&=&A\;u^{\mu}u^{\nu}u^{\rho}u^{\sigma}\nn
&+&B\left[u^{\mu}u^{\nu}\Delta^{\rho\sigma}+u^{\mu}u^{\rho}\Delta^{\nu\sigma}+u^{\mu}u^{\sigma}\Delta^{\nu\rho}+u^{\rho}u^{\nu}\Delta^{\mu\sigma}+u^{\sigma}u^{\nu}\Delta^{\mu\rho}+u^{\rho}u^{\sigma}\Delta^{\nu\mu}\right]\nn
&+&C\left[u^{\mu}u^{\nu}\Pi^{\rho\sigma}+u^{\mu}u^{\rho}\Pi^{\nu\sigma}+u^{\mu}u^{\sigma}\Pi^{\nu\rho}+u^{\rho}u^{\nu}\Pi^{\mu\sigma}+u^{\sigma}u^{\nu}\Pi^{\mu\rho}+u^{\rho}u^{\sigma}\Pi^{\nu\mu}\right]\nn
&+&D\left[\Delta^{\nu\mu}\Delta^{\rho\sigma}+\Delta^{\mu\rho}\Delta^{\nu\sigma}+\Delta^{\mu\sigma}\Delta^{\nu\rho}\right]\nn
&+&E\left[\Delta^{\nu\mu}\Pi^{\rho\sigma}+\Delta^{\mu\rho}\Pi^{\nu\sigma}+\Delta^{\mu\sigma}\Pi^{\nu\rho}+\Delta^{\nu\rho}\Pi^{\mu\sigma}+\Delta^{\nu\sigma}\Pi^{\mu\rho}+\Delta^{\rho\sigma}\Pi^{\nu\mu}\right].
\tea
The condition that $u_{\mu}K^{\mu\nu\rho\sigma}=-A^{\nu\rho\sigma}$ reduces this to
\bea
K^{\mu\nu\rho\sigma}&=&\rho\;u^{\mu}u^{\nu}u^{\rho}u^{\sigma}\nn
&+&\frac13\rho\left[u^{\mu}u^{\nu}\Delta^{\rho\sigma}+u^{\mu}u^{\rho}\Delta^{\nu\sigma}+u^{\mu}u^{\sigma}\Delta^{\nu\rho}+u^{\rho}u^{\nu}\Delta^{\mu\sigma}+u^{\sigma}u^{\nu}\Delta^{\mu\rho}+u^{\rho}u^{\sigma}\Delta^{\nu\mu}\right]\nn
&+&\rho\left[u^{\mu}u^{\nu}\Pi^{\rho\sigma}+u^{\mu}u^{\rho}\Pi^{\nu\sigma}+u^{\mu}u^{\sigma}\Pi^{\nu\rho}+u^{\rho}u^{\nu}\Pi^{\mu\sigma}+u^{\sigma}u^{\nu}\Pi^{\mu\rho}+u^{\rho}u^{\sigma}\Pi^{\nu\mu}\right]\nn
&+&D\left[\Delta^{\nu\mu}\Delta^{\rho\sigma}+\Delta^{\mu\rho}\Delta^{\nu\sigma}+\Delta^{\mu\sigma}\Delta^{\nu\rho}\right]\nn
&+&E\left[\Delta^{\nu\mu}\Pi^{\rho\sigma}+\Delta^{\mu\rho}\Pi^{\nu\sigma}+\Delta^{\mu\sigma}\Pi^{\nu\rho}+\Delta^{\nu\rho}\Pi^{\mu\sigma}+\Delta^{\nu\sigma}\Pi^{\mu\rho}+\Delta^{\rho\sigma}\Pi^{\nu\mu}\right].\tea
Tracelessnes implies
\bea
0&=&-\frac13\rho\Delta^{\rho\sigma}+5D\Delta^{\rho\sigma}\nn
0&=&-\rho+7E.
\tea
We therefore may write
\bea
K^{\mu\nu\rho\sigma}&=&\rho\left\{\;u^{\mu}u^{\nu}u^{\rho}u^{\sigma}\right.\nn
&+&\frac13\left[u^{\mu}u^{\nu}\Delta^{\rho\sigma}+u^{\mu}u^{\rho}\Delta^{\nu\sigma}+u^{\mu}u^{\sigma}\Delta^{\nu\rho}+u^{\rho}u^{\nu}\Delta^{\mu\sigma}+u^{\sigma}u^{\nu}\Delta^{\mu\rho}+u^{\rho}u^{\sigma}\Delta^{\nu\mu}\right]\nn
&+&u^{\mu}u^{\nu}\Pi^{\rho\sigma}+u^{\mu}u^{\rho}\Pi^{\nu\sigma}+u^{\mu}u^{\sigma}\Pi^{\nu\rho}+u^{\rho}u^{\nu}\Pi^{\mu\sigma}+u^{\sigma}u^{\nu}\Pi^{\mu\rho}+u^{\rho}u^{\sigma}\Pi^{\nu\mu}\nn
&+&\frac1{15}\left[\Delta^{\nu\mu}\Delta^{\rho\sigma}+\Delta^{\mu\rho}\Delta^{\nu\sigma}+\Delta^{\mu\sigma}\Delta^{\nu\rho}\right]\nn
&+&\left.\frac17\left[\Delta^{\nu\mu}\Pi^{\rho\sigma}+\Delta^{\mu\rho}\Pi^{\nu\sigma}+\Delta^{\mu\sigma}\Pi^{\nu\rho}+\Delta^{\nu\rho}\Pi^{\mu\sigma}+\Delta^{\nu\sigma}\Pi^{\mu\rho}+\Delta^{\rho\sigma}\Pi^{\nu\mu}\right]\right\}.
\tea
This contributes to the equations of motion a term
\bea
&&S^{\mu\nu}_{\lambda\chi}K^{\lambda\chi\rho\sigma}u_{\rho,\sigma}=\rho\left\{\frac1{15}\left[\Delta^{\nu\sigma}u^{\mu}_{,\sigma}+\Delta^{\mu\sigma}u^{\nu}_{,\sigma}-\frac23\Delta^{\mu\nu}\theta\right]\right.\nn
&+&\left.\frac17\left[\Pi^{\nu\sigma}u^{\mu}_{,\sigma}+\Pi^{\mu\sigma}u^{\nu}_{,\sigma}+\Delta^{\mu\sigma}\Pi^{\nu\rho}u_{\rho,\sigma}+\Delta^{\nu\sigma}\Pi^{\mu\rho}u_{\rho,\sigma}+\Pi^{\nu\mu}\theta-\frac43\Delta^{\nu\mu}\Pi^{\rho\sigma}u_{\rho,\sigma}\right]\right\}.
\tea
We thus obtain eq. (\ref{MC}).

\section{Momentum monomials}
\label{B}
In computing momentum integrals in the text we have used the identities
\bea
\left.\frac{p^j}{p}\right|_{p=0}&=&0\nn
\left.\frac{p^{j}p^{k}}{p^2}\right|_{p=0}&=&\frac13\delta^{jk},
\tea
leading to 
\bea
\left.\frac{\left(pq\right)}{pq}\right|_{p=0}&=&0\nn
\left.\frac{\left(pq\right)^2}{p^2q^2}\right|_{p=0}&=&\frac13.
\tea
Higher monomials were computed from Wick's theorem
\bea
\left.\frac{\left(pq\right)^{2k+1}}{p^{2k+1}q^{2k+1}}\right|_{p=0}&=&0\nn
\left.\frac{\left(pq\right)^{2k}}{p^{2k}q^{2k}}\right|_{p=0}&=&\frac1{2k+1}.
\tea

\section*{Acknowledgments}
I thank A. Kandus, N. Mir\'on Granese, L. Cantarutti, M. Nigro and G. E. Perna for multiple discussions.
This work  was supported in part by Universidad de Buenos Aires through
grant UBACYT 20020170100129BA, CONICET and ANPCyT.


\begin{thebibliography}{99}

\bibitem{BN11} A. Brandenburg and A. Nordlund, 
\textit{Astrophysical turbulence modeling},
Rept. Prog. Phys. 74, 046901 (2011). 

\bibitem{LBM13} P. D. Lasky, M. F. Bennett and A. Melatos,
\textit{Stochastic gravitational wave background from hydrodynamic turbulence in
differentially rotating neutron stars},
Phys. Rev. D 87, 063004   (2013).


\bibitem{RZ13} L. Rezzolla and O. Zanotti, 
\textit{Relativistic Hydrodynamics} 
(Oxford University Press, Oxford, 2013).

\bibitem{H12} U. Heinz,
\textit{Towards the Little Bang Standard Model},
Journal of Physics: Conference Series 455, 012044 (2013).

\bibitem{RR19} P. Romatschke and U. Romatschke, 
\textit{Relativistic fluid dynamics in and out equilibrium}, 
Cambridge University Press (2019).

\bibitem{L19} K. Lozanov,
\textit{Lectures on Reheating after Inflation},
arXiv:1907.04402 (2019).


\bibitem{FFTV20} D. G. Figueroa, A. Florio, F. Torrenti, W. Valkenburg, 
\textit{The art of simulating the early Universe -- Part I}, 
arXiv:2006.15122 (2020).

\bibitem{MT04} R. Micha and I. Tkachev,
\textit{Turbulent thermalization},
Phys. Rev. D 70, 043538 (2004).

\bibitem{BBSV14} J. Berges, K. Boguslavski, S. Schlichting, and R. Venugopalan, 
\textit{Turbulent Thermalization Process in Heavy Ion Collisions at Ultrarelativistic Energies},
Phys. Rev. D 89, 074011 (2014).

\bibitem{CH08} E. Calzetta and B-L. Hu,
\textit{Nonequilibrium quantum field theory},
(Cambridge University Press, London (2008)).

\bibitem{Bazow16} D. Bazow, G. S. Denicol, U. Heinz, M. Martinez and J. Noronha,
\textit{Nonlinear dynamics from the relativistic Boltzmann equation in the
Friedmann-Lema\^\i tre-Robertson-Walker spacetime},
Phys. Rev. D 94, 125006 (2016).

\bibitem{SND} M. Strickland, J. Noronha and G. Denicol,  
\textit{Anisotropic nonequilibrium hydrodynamic attractor}, 
Phys. Rev. D 97, 036020  (2018).


 

\bibitem{MT03} R. Micha and I. Tkachev,
\textit{Relativistic Turbulence: A Long Way from Preheating to Equilibrium},
Phys. Rev. Lett. 90,  121301 (2003).

\bibitem{GC02}M. Gra\~na and E. Calzetta, 
\textit{Reheating and Turbulence}, 
Phys. Rev. D 65, 063522 (2002).

\bibitem{K08}V. Khachatryan,
\textit{Modified Kolmogorov Wave Turbulence in QCD matched onto Bottom-up Thermalization},
Nucl. Phys. A810, 109 (2008).

\bibitem{FW11} Stefan Floerchinger and Urs Achim Wiedemann,
\textit{Fluctuations around Bjorken flow and the onset of turbulent phenomena},
JHEP 11, 100 (2011).

\bibitem{CR11} M. E. Carrington and A. Rheban,
\textit{Perturbative and Nonperturbative Kolmogorov Turbulence in a Gluon Plasma},
The European Physical Journal C 71, 1787 (2011).

\bibitem{F13} K. Fukushima,
\textit{Turbulent pattern formation and diffusion in the early-time dynamics in the relativistic heavy-ion collision},
Phys.Rev. C89, 024907 (2014). 

\bibitem{AKLN14} M. C. Abraao York, A. Kurkela, E. Lu and G. D. Moore,
\textit{UV Cascade in Classical Yang-Mills via Kinetic Theory},
Phys. Rev. D 89, 074036 (2014).


\bibitem{ED18} G. L. Eyink and Th. D. Drivas,
\textit{Cascades and Dissipative Anomalies in Relativistic Fluid Turbulence},
Phys. Rev. X 8, 011023 (2018).

\bibitem{O49} A. M. Obukhov, 
\textit{Temperature Field Structure in a Turbulent Flow}, 
Izvestiia Akademii Nauk SSSR, Ser. Geogr. i Geofiz 13, 58 (1949).

\bibitem{ED18b} G. L. Eyink and Th. D. Drivas,
\textit{Cascades and Dissipative Anomalies in Compressible Fluid Turbulence}, 
Phys. Rev. X 8, 011022 (2018).

\bibitem{E18} G. L. Eyink,
\textit{Cascades and Dissipative Anomalies in Nearly Collisionless Plasma Turbulence},
Phys. Rev. X 8, 041020 (2018).


\bibitem{Biferale} A. Alexakis and L. Biferale,
\textit{Cascades and transitions in turbulent flows},  
Physics Reports 767-769,  1 (2018).

\bibitem{RR13} D. Radice and L. Rezzolla,
\textit{Universality and intermittency in relativistic turbulent flows of a hot plasma},
Astrophys. J., 766, L10 (2013).

\bibitem{ZR13} J. Zrake and A. I. MacFadyen,
\textit{Spectral and intermittency properties of relativistic turbulence},
Astrophys. J., 763, L12, (2013).

\bibitem{Lehner12} F. Carrasco, L. Lehner, R. C. Myers, O. Reula and A. Singh,
\textit{Turbulent flows for relativistic conformal fluids in 2+1 dimensions}
Physical Review D 86,  126006 (2012).

\bibitem{Oz15} J. R. Westernacher-Schneider, L. Lehner and Y. Oz,
\textit{Scaling Relations in Two-Dimensional Relativistic Hydrodynamic Turbulence},
JHEP 2015, 67 (2015).



\bibitem{VanBiro12} P. V\'an and T. S. Bir\'o, 
\textit{First order and stable relativistic dissipative hydrodynamics}, 
Phys. Lett. B 709,  106 (2012).


\bibitem{BDN18} F. S. Bemfica, M. M. Disconzi, and J. Noronha,
\textit{Causality and existence of solutions of relativistic viscous fluid dynamics with gravity},
Phys.Rev.D 98, 104064  (2018). 

\bibitem{BDN19} F. S. Bemfica, M. M. Disconzi, and J. Noronha,
\textit{Nonlinear causality of general first-order relativistic viscous hydrodynamics},
Phys.Rev.D 100, 104020 (2019).

\bibitem{BDN20} F. S. Bemfica, M. M. Disconzi, and J. Noronha,
\textit{General-Relativistic Viscous Fluid Dynamics}, 
arXiv:2009.11388 (2020).


 
\bibitem{kovtun19} P. Kovtun, 
\textit{First-order relativistic hydrodynamics is stable}, 
JHEP {10},  034 (2019).

\bibitem{DasFlorNoRy20} A. Das, W. Florkowski, J. Noronha and R. Ryblewski, 
\textit{Equivalence between first-order causal and stable hydrodynamics and Israel-Stewart theory for boost-invariant systems with a constant relaxation time}, 
Phys. Lett. B 806,  135525 (2020).

\bibitem{GPRuRe19} A. L. Garc\'\i a-Preciante, M. E. Rubio and O. A. Reula, 
\textit{Generic instabilities in the relativistic Chapman-Enskog heat conduction law}, 
J Stat Phys 181, 246 (2020). 

\bibitem{HK20}R. E. Hoult and  P. Kovtun, 
\textit{Stable and causal relativistic Navier-Stokes equations},  
JHEP 06  067 (2020).

\bibitem{FO10} I. Fouxon and Y. Oz,
\textit{Exact scaling relations in relativistic hydrodynamic turbulence},
Phys. Lett. B 694, 261 (2010).

\bibitem{israel76} W. Israel,  
\textit{Nonstationary irreversible thermodynamics: A causal relativistic theory},
Ann. Phys. (NY) {100}, 310 (1976).

\bibitem{IsSte76} W. Israel and J. M. Stewart, 
\textit{Thermodynamics of nonstationary and transient effects in a relativistic gas},
Phys. Lett. A {58}, 213 (1976).

\bibitem{IsSte79a} W. Israel and M. Stewart, 
\textit{Transient relativistic thermodynamics and kinetic theory},
Ann. Phys. (NY) {118}, 341 (1979).

\bibitem{IsSte79b} W. Israel and M. Stewart, 
\textit{On transient relativistic thermodynamics and kinetic theory. II},
Proc. R. Soc. London, Ser A {365}, 43 (1979).

\bibitem{IsSte80} W. Israel and J. M. Stewart, 
\textit{Progress in relativistic thermodynamics and electrodynamics of continuous media}, 
{General Relativity and Gravitation} 2, 491 (1980).

\bibitem{LiMuRu86} I. S. Liu, I. M\"uller and T. Ruggeri, 
\textit{Relativistic thermodynamics of gases},
Ann. Phys. {169}, 191 (1986).

\bibitem{GerLind90} R. Geroch and L. Lindblom, 
\textit{Dissipative relativistic fluid theories of divergence type},
Phys. Rev. D {41},  1855 (1990). 

\bibitem{GerLind91} R. Geroch and L. Lindblom, 
\textit{Causal theories of dissipative relativistic fluids}
Ann. Phys. (NY) 207, 394 (1991).

\bibitem{ReNa97} O. A. Reula and G. B. Nagy, 
\textit{On the causality of a dilute gas as a dissipative relativistic fluid
theory of divergence type}, 
J. Phys.  {A 28}, 6943 (1995).

\bibitem{PRCal09} J. Peralta-Ramos and E. Calzetta, 
\textit{Divergence-type nonlinear conformal hydrodynamics},
Phys. Rev. D {80}, 126002 (2009).

\bibitem{PRCal10} J. Peralta-Ramos and E. Calzetta, 
\textit{Divergence-type 2+1 dissipative hydrodynamics applied to heavy-ion collisions}
Phys. Rev. C {82},  054905 (2010).

\bibitem{cal15} E. Calzetta, 
\textit{Hydrodynamic approach to boost invariant free-streaming},
Phys. Rev. D {92}, 045035 (2015).

\bibitem{cal98} E. Calzetta,  
\textit{Relativistic fluctuating hydrodynamics},
 Class. Quant. Grav. {15},  653 (1998).

\bibitem{LheReRu18} L. Lehner, O. A. Reula and M. E. Rubio,  
\textit{A Hyperbolic Theory of Relativistic Conformal Dissipative Fluids},
Phys. Rev. D {97},  024013 (2018).

\bibitem{lucas19} L. Cantarutti and E. Calzetta, 
\textit{Dissipative-type theories for Bjorken and Gubser flows}, 
Int. J. Mod. Phys. A {35}, 2050074 (2020).

\bibitem{Nahuel20} N. Mir\'on-Granese, A. Kandus and E. Calzetta, 
\textit{Nonlinear fluctuations in relativistic causal fluids}, 
JHEP 07, 064 (2020).

\bibitem{DMNR10} G.S. Denicol, T. Koide, and D.H. Rischke, 
\textit{Dissipative relativistic fluid dynamics: a new way to derive the equations of motion from kinetic theory}, 
Phys.Rev.Lett. 105, 162501 (2010).

\bibitem{DMNR11} B. Betz, G.S. Denicol, T. Koide, E. Moln\'ar, H. Niemi, and D.H. Rischke, 
\textit{Second order dissipative fluid dynamics from kinetic theory}, 
Eur.Phys.J. Conf. 13, 07005 (2011).

\bibitem{DMNR11b} G.S. Denicol, J. Noronha, H. Niemi, and D.H. Rischke,   
\textit{Origin of the relaxation time in dissipative fluid dynamics}, 
{Phys. Rev.} D 83, 074019 (2011).

\bibitem{DMNR12}  G. S. Denicol, E. Moln\'ar, H. Niemi and D. H. Rischke, 
\textit{Derivation of fluid dynamics from kinetic theory with the 14 moment approximation}, 
Eur. Phys. J. A 48 11 (2012).

\bibitem{DMNR12b} G. S. Denicol, H. Niemi, E. Moln\'ar and D. H. Rischke
\textit{Derivation of transient relativistic fluid dynamics from the Boltzmann equation},
{Phys. Rev.} D 85, 114047 (2012); (E) {Phys. Rev.} D 91, 039902(E) (2015).

\bibitem{DMNR14} G. S. Denicol, H. Niemi, I. Bouras, E. Moln\'ar, Z. Xue, D. H. Rischke, and C. Greiner, 
\textit{Solving the heat-flow problem with transient relativistic fluid dynamics}, 
Phys.Rev. D89, 074005 (2014).

\bibitem{DMNR12c} G. S. Denicol and H. Niemi, 
\textit{Derivation of transient relativistic fluid dynamics from the Boltzmann equation for a multi-component system}, 
Nucl. Phys. A 904-905, 369 (2013).

\bibitem{DMNR14b} E. Moln\'ar,  H. Niemi, G. S. Denicol, and D. H. Rischke,   
\textit{Relative importance of second-order terms in relativistic dissipative fluid dynamics}, 
{Phys. Rev.} D 89, 074010 (2014).

\bibitem{DMNR16} H. Niemi and G. S. Denicol,   
\textit{How large is the Knudsen number reached in fluid dynamical simulations of ultrarelativistic heavy ion collisions?}, 
arXiv:1404.7327 (2014).

\bibitem{ST14} M. Strickland, 
\textit{Anisotropic Hydrodynamics: Three lectures}, 
Act. Phys. Pol. B 45, 2355 (2014).

\bibitem{ST14b} M. Strickland,
\textit{Anisotropic Hydrodynamics: Motivation and Methodology},
Nucl. Phys. A 926, 92 (2014).

\bibitem{FMR15} W. Florkowski, E. Maksymiuk, R. Ryblewski, L. Tinti, 
\textit{Anisotropic hydrodynamics for mixture of quark and gluon fluids}, 
{Phys. Rev.} C 92, 054912 (2015).

\bibitem{FRST16} W. Florkowski, R. Ryblewski, M. Strickland, L. Tinti, 
\textit{Non-boost-invariant dissipative hydrodynamics}, 
{Phys. Rev.} C 94, 064903 (2016).

\bibitem{NMR17} H. Niemi, E. Moln\'ar, and D. H. Rischke,   
\textit{The right choice of moment for anisotropic fluid dynamics}, 
Nucl.Phys. A967, 409  (2017).


\bibitem{BHS14} D. Bazow, U. Heinz, M. Strickland, 
\textit{Second-order (2+1)-dimensional anisotropic hydrodynamics},
Phys. Rev. C 90, 054910 (2014).

\bibitem{BNR16a} E. Moln\'ar, H. Niemi, D. H. Rischke, 
\textit{Derivation of anisotropic dissipative fluid dynamics from the Boltzmann equation}, 
{Phys. Rev.} D 93, 114025 (2016).

\bibitem{BNR16b} E. Moln\'ar, H. Niemi, D. H. Rischke, 
\textit{Closing the equations of motion of anisotropic fluid dynamics by a judicious choice of a moment of the Boltzmann equation}, 
{Phys. Rev.} D 94, 125003 (2016).

\bibitem{NBH21} M. McNelis, D. Bazow and  U. Heinz,
\textit{Anisotropic fluid dynamical simulations of heavy-ion collisions},
arXiv::2101.02827 (2021).

\bibitem{MGC17} N. Mir\'on Granese and E. Calzetta, 
\textit{Primordial gravitational waves amplification from causal fluids},
Phys. Rev. D {97}, 023517  (2018).

\bibitem{MG21}  N. Mir\'on Granese,
\textit{Relativistic viscous effects on the primordial gravitational waves spectrum},
arXiv:2012.11422 (2020).

\bibitem{CF} C. Caprini and D.G. Figueroa, 
\textit{Cosmological Backgrounds of Gravitational Waves}, 
Class. Quant. Grav. 35, 163001  (2018).


\bibitem{Max67} J. C. Maxwell, 
\textit{On the dynamical theory of gases},
Philos. Trans. Soc. London {157},  49 (1867).

\bibitem{Catt48} C. Cattaneo,  
\textit{Sulla conduzione del calore},
Atti del Semin. Mat. e Fis. Univ. Modena {3}, 3  (1948).

\bibitem{Catt58} C. Cattaneo, 
\textit{Sur la propagation de la chaleur en relativit\'e},
C. R. Acad. Sci. Paris {247},  431 (1958).

\bibitem{JosPrez89} D. D. Joseph and L. Preziosi, 
\textit{Heat waves},
Rev. Mod. Phys. {61}, 41 (1989); Addendum, Rev. Mod. Phys. {62}, 375 (1990).




\bibitem{PRC10a}E. Calzetta and J. Peralta-Ramos,
\textit{Linking the hydrodynamic and kinetic description of a dissipative relativistic conformal theory},
{Phys. Rev.} D82, 106003 (2010).

\bibitem{PRC13a}J. Peralta-Ramos and E. Calzetta,
\textit{Macroscopic approximation to relativistic kinetic theory from a nonlinear closure},
{Phys. Rev.} D 87, 034003 (2013).

\bibitem{vKH}Th. de Karman and L. Howarth,
\textit{On the statistical theory of isotropic turbulence}, 
Proc. R. Soc. Lond. A  164, 192 (1938).

\bibitem{LL87} L.D. Landau and E.M. Lifshitz, 
\textit{Fluid Mechanics}, 2nd ed. 
(Pergamon Press, Oxford (1987)).

\bibitem{Frisch} U. Frisch, 
\textit{Turbulence} 
(Cambridge U. P., (1995))

\bibitem{MY71} A. S. Monin and  A. M. Yaglom,  
\textit{Statistical Fluid Mechanics}, vol. 1,
ed. J. Lumley. (MIT Press, Cambridge, MA., (1971)).

\bibitem{MY75}A. S. Monin and  A. M. Yaglom,  
\textit{Statistical Fluid Mechanics}, vol. 2,
ed. J. Lumley. (MIT Press, Cambridge, MA., (1975)).

\bibitem{Chandra54} E. A. Spiegel (Editor),
\textit{The Theory of Turbulence: Subrahmanyan Chandrasekhar's 1954 Lectures}
Lect. Notes Phys. 810 (Springer, Dordrecht 2011).

\bibitem{Pope} S. B. Pope,
\textit{Turbulent Flows}
(Cambridge U. P. (2000)).


\bibitem{Muller99} I. M\"uller, 
\textit{Speeds of Propagation in Classical and Relativistic Extended Thermodynamics}, 
Living Rev. Relativity, 2, 1 (1999).

\bibitem{BL99} G. Boillat and T. Ruggeri, 
\textit{Relativistic gas: Moment equations and maximum wave velocity}, 
J. Math. Phys. 40, 6399 (1999).


\bibitem{BRS08} J. Berges, A. Rothkopf, and J. Schmidt, 
\textit{Nonthermal Fixed Points: Effective Weak Coupling for Strongly Correlated Systems Far from Equilibrium},
Phys. Rev. Lett. 101, 041603 (2008).

\bibitem{Berges15} A. Pi\~neiro Orioli, K. Boguslavski and J. Berges,
\textit{Universal self-similar dynamics of relativistic and nonrelativistic field theories near
nonthermal fixed points},
Phys. Rev. D 92, 025041 (2015).

\bibitem{NEKSSG12} B. Nowak, S. Erne, M. Karl, J. Schole, D. Sexty, Th. Gasenzer, 
\textit{Non-thermal fixed points: universality, topology, and turbulence in Bose gases}, 
Proceedings of the Summer school: Strongly interacting quantum systems out of equilibrium,  Les Houches, France (2012).

\bibitem{SMG19} Ch-M. Schmied, A. N. Mikheev, Th. Gasenzer, 
\textit{Non-thermal fixed points: Universal dynamics far from equilibrium}, 
Int. J. Mod. Phys. A 34, 29 (2019).





\bibitem{AW74a} J. L. Anderson and H. R. Witting, 
\textit{A relativistic relaxation-time model for the Boltzmann equation}, 
Physica (Amsterdam) 74,  466 (1974).

\bibitem{AW74b} J. L. Anderson and H. R. Witting, 
\textit{Relativistic quantum transport coefficients},
Physica (Amsterdam) 74, 489 (1974).


\end{thebibliography}
\end{document}